\newif\iffigs\figstrue
\documentclass[12pt]{article}
\iffigs
\input{epsf}
\else
\fi

\newcommand{\eqn}[1]{(\ref{#1})}

\newsavebox{\uuunit}
\sbox{\uuunit}
{\setlength{\unitlength}{0.825em}
\begin{picture}(0.6,0.7)
\thinlines
\put(0,0){\line(1,0){0.5}}
\put(0.15,0){\line(0,1){0.7}}
\put(0.35,0){\line(0,1){0.8}}
\multiput(0.3,0.8)(-0.04,-0.02){12}{\rule{0.5pt}{0.5pt}}
\end {picture}}

\def\IP{\relax{\rm I\kern-.18em P}}

\begin{document}
%
\font\cmss=cmss10 \font\cmsss=cmss10 at 7pt
\def\twomat#1#2#3#4{\left(\matrix{#1 & #2 \cr #3 & #4}\right)}
\def\inbar{\vrule height1.5ex width.4pt depth0pt}
\def\IC{\relax\,\hbox{$\inbar\kern-.3em{\rm C}$}}
\def\IG{\relax\,\hbox{$\inbar\kern-.3em{\rm G}$}}
\def\IB{\relax{\rm I\kern-.18em B}}
\def\ID{\relax{\rm I\kern-.18em D}}
\def\IL{\relax{\rm I\kern-.18em L}}
\def\IF{\relax{\rm I\kern-.18em F}}
\def\IH{\relax{\rm I\kern-.18em H}}
\def\II{\relax{\rm I\kern-.17em I}}
\def\IN{\relax{\rm I\kern-.18em N}}
\def\IP{\relax{\rm I\kern-.18em P}}
\def\IQ{\relax\,\hbox{$\inbar\kern-.3em{\rm Q}$}}
\def\bfzero{\relax\,\hbox{$\inbar\kern-.3em{\rm 0}$}}
\def\IK{\relax{\rm I\kern-.18em K}}
\def\IG{\relax\,\hbox{$\inbar\kern-.3em{\rm G}$}}
 \font\cmss=cmss10 \font\cmsss=cmss10 at 7pt
\def\IR{\relax{\rm I\kern-.18em R}}
\def\ZZ{\relax\ifmmode\mathchoice
{\hbox{\cmss Z\kern-.4em Z}}{\hbox{\cmss Z\kern-.4em Z}}
{\lower.9pt\hbox{\cmsss Z\kern-.4em Z}}
{\lower1.2pt\hbox{\cmsss Z\kern-.4em Z}}\else{\cmss Z\kern-.4em
Z}\fi}
\def\bfone{\relax{\rm 1\kern-.35em 1}}
\def\dop{{\rm d}\hskip -1pt}
\def\real{{\rm Re}\hskip 1pt}
\def\trace{{\rm Tr}\hskip 1pt}
\def\ii{{\rm i}}
\def\diag{{\rm diag}}
\def\sch#1#2{\{#1;#2\}}
\def\nn{\nonumber}
\def\spa{&&\!\!\!\!\!\!\!\!\!\!\!\!}
\begin{titlepage}
\setcounter{page}{0}
\vskip 40pt
\begin{flushright}
SISSA REF 26/98/EP
\end{flushright}

\vskip 80pt

\begin{center}
{\Large \bf Black holes as D3--branes on Calabi--Yau threefolds}

\vskip 20pt

{Matteo Bertolini$^a$, Pietro Fr\`e$^b$, Roberto Iengo$^a$ and Claudio 
A. Scrucca$^a$}

\vskip 20pt

{\it $^a$International School for Advanced Studies ISAS--SISSA and INFN \\
Sezione di Trieste, Via Beirut 2--4, 34013 Trieste, Italy}

\vskip 5pt

{\it $^b$Dipartimento di Fisica Teorica, Universit\`a di Torino and INFN \\
Sezione di Torino, Via P. Giuria 1, 10125 Torino, Italy}

\vskip 60pt

\end{center}

\begin{abstract}

\vskip 20pt

We show how an extremal Reissner--Nordstr\"om black hole can be obtained
by wrapping a dyonic D3--brane on a Calabi--Yau manifold. In the orbifold
limit $T^6/ \ZZ_3$, we explicitly show the correspondence between the
solution of the supergravity equations of motion and the D--brane boundary
state description of such a black hole.

\end{abstract}

\vskip 60pt

\begin{flushleft}



\end{flushleft}

\vspace{2mm} \vfill \hrule width 3.cm
\vskip 0.2cm
{\footnotesize 
Supported in part by   EEC  under TMR contract
ERBFMRX--CT96--0045, in which M. Bertolini, R. Iengo and C. A. Scrucca
are associated to Frascati}.
\end{titlepage}
\newpage

In the last couple of years there has been much effort in finding a
microscopic description of both extremal and non--extremal
black holes arising as compactifications of different {\it p}--brane
solutions of ten--dimensional supergravity theories. This has been done
by considering various solitonic configurations in string theory, such
as bound states of D--branes and solitons of different kinds \cite{mal}
or as  intersecting (both orthogonally and at angles) D--branes alone
\cite{bal}.
As far as the microscopic description is concerned, these studies have
been mainly devoted to toroidal compactifications and less has been said
about Calabi--Yau (CY) ones. On the contrary, from a macroscopic (i.e. supergravity)
point of view, these black hole solutions have been known for a long time
in both cases and many progresses have been made in the last few years
(see \cite{fer} and many subsequent  works). Different problems arise when
trying to find an appropriate D--brane description of these solutions in a
non--flat asymptotic space. Moreover, some general results that are valid
in the toroidal case no longer hold for CY compactifications.
In particular, it is not straightforward to generalize the so called
``harmonic function rule'' and it is also no longer true that the minimum
number of ``different'' charges (that is, carried by different microscopic
objects) must be $4$ in order to obtain a regular black hole in four
dimensions. 

We will be interested in discussing a Reissner--Nordstr\"om (R--N) black hole in four 
dimensions within a CY compactification (whose relevance for obtaining non--singular 
four--dimensional black hole was already pointed out, see for instance 
ref. \cite{lust1}). The R--N solution defined as the usual non-singular black 
hole solution of Maxwell--Einstein gravity, can also be seen as a 
particular solution of a wider class of field theories in four dimensions in which 
the only fields having a non--trivial coordinate dependence are the metric 
$G_{\mu\nu}$ and a gauge field $A_{\mu}$, whereas any other field is
taken to be constant. In particular, in four--dimensional N=2 supergravity
this solution, known as the {\it double--extreme} black--hole \cite{kal},
arises in the specific case in which one assumes that the moduli fields
belonging to vector multiplets (as well as those belonging to the
hyper-multiplets which are anyhow constant in any N=2 black--hole solution)
take the same constant values from the horizon to spatial infinity. In order
to be consistent with the field equations such constant values are not arbitrary
but must coincide with the so called {\it fixed values}: these are
determined in terms of the electric and magnetic charges of all the
existing gauge fields by a variational principle that extremizes the
central charge and leads to classical formulae expressing the
horizon area as a quartic invariant of the U--duality group
(see for instance \cite{frelec,ferraralec,sabra} and references therein).

When ten--dimensional supergravity is compactified on a CY
threefold  ${\cal M}^{CY}_3$ we obtain $D=4,N=2$ supergravity coupled to matter.
As well known the field content of the four--dimensional theory
and its interaction structure is completely determined by the
{\it topological and analytical type} of  ${\cal M}^{CY}_3$
but depends in no way on its metric structure.
Indeed the standard counting of hyper and vector multiplets tells us that 
$n_V=h^{(1,2)}$ and $n_H=h^{(1,1)}+1$, the numbers $h^{(p,q)}$ being the 
dimensions of the Dolbeault cohomology groups.
Furthermore, the geometrical datum that completely specifies the
vector multiplet coupling, namely the choice of the special K\"ahler
manifold and its special K\"ahler metric is provided by the moduli
space geometry of complex structure deformations. To determine this
latter no reference has ever to be made to the K\"ahler
metric $g_{ij^\star}$ installed on ${\cal M}^{CY}_3$ (for a review of this 
well established results see for instance \cite{fresoriani}).
Because of this crucial property careful thought is therefore needed when one tries 
to {\it oxidize} the solutions of  four--dimensional $N=2$ supergravity obtained 
through compactification on ${\cal M}^{CY}_3$ to {\it bona fide} solutions
of the original ten--dimensional Type IIB supergravity.
To see the four--dimensional configuration as a configuration in ten--dimension 
one has to choose a metric on the internal manifold in such a way as to satisfy 
the full set of ten--dimensional equations.

\vskip 7pt

In this note we will show how an extreme R--N black hole can be
obtained by compactification of the self--dual D3--brane on 
${\cal M}^{CY}_3=T^6/ \ZZ_3$, which is the orbifold limit of a CY manifold with 
Hodge numbers $h^{(1,1)}=9$ and $h^{(1,2)}=0$. Recalling some results obtained in
previous works \cite{hins1,bis}, we will explicitly show the correspondence between 
the supergravity solution and the D--brane boundary state description of such a
black hole. In this case, the effective four--dimensional theory is N=2 supergravity
coupled to $10$ hypermultiplets and $0$ vector multiplets, the only vector field
in the game being the graviphoton. Since there are no vector multiplet scalars the 
only regular black hole solution is the double--extreme one. From a supergravity point 
of view this is somewhat obvious and the same conclusion  holds for every Type IIB 
compactification on CY manifolds with $h^{(1,2)}=0$. The interest of the $T^6/ \ZZ_3$ 
case lies in the fact that an explicit and simple D--brane boundary state description 
can be found. It would be obviously very interesting to find more complicated configurations
which correspond to regular $N=2$ black hole solutions for which an analogous D--brane
description can be found.

We will start by showing that the {\it oxidization} of a {\it double extreme}
black--hole solution of $N=2$ supergravity to a {\it bona fide}
solution of Type IIB supergravity is possible and quite
straightforward. It just suffices to choose for the CY
metric the Ricci flat one whose existence in every K\"ahler class is
guaranteed by Yau theorem \cite{Yau}. Our exact solution of Type IIB
supergravity in ten dimensions corresponds to a 3--brane wrapped on a 3--cycle of 
the generic threefold ${\cal M}^{CY}_3$ and dimensionally reduced to 4--dimensions
is a double--extreme black hole.  Let us then argue how this simple
result is obtained.

As well known, prior to the recent work by Bandos, Sorokin and Tonin 
\cite{BST} Type IIB supergravity had no supersymmetric space--time action. Only 
the field equations could be written as closure conditions of the supersymmetry 
algebra \cite{schwarz1}. The same result could be obtained from the rheonomy
superspace formalism as shown in \cite{leopesa}. Indeed, the condition
of self--duality for the R--R 5--form $F_{(5)}$ that is necessary for the equality 
of Bose and Fermi degrees of freedom  cannot be easily obtained as a variational 
equation and has to be stated as a constraint. In the
new approach of \cite{BST} such problems are circumvented by
introducing more fields and more symmetries that remove spurious
degrees of freedom. For our purposes these
subtleties are not relevant since our goal is that of showing the
existence of a classical solution. Hence we just need the field equations
which are unambiguous and reduce, with our ansatz, to the following
ones:
\begin{eqnarray}
\label{E10}
R_{MN} \!\!\!\! &=& \!\!\!\! T_{MN} \\
\nabla _M F_{(5)}^{MABCD}=0 \,
\; & \longleftarrow & \;  F^{(5)}_{G_1 \dots G_5}
= \frac{1}{5!} \, \epsilon_{G_1 \dots G_5 H_1 \dots H_5}\,F_{(5)}^{H_1 \dots H_5}
\end{eqnarray}
$T_{MN}= 1/(2 \cdot 4!) \, F^{(5)}_{M \cdot \cdot \cdot\cdot } \,
F^{(5)}_{N \cdot \cdot \cdot \cdot }$ being the traceless energy--momentum tensor
of the R--R 4--form $A_{(4)}$ to which the 3--brane couples and
$F_{(5)}$   the corresponding self--dual field strength.

It is noteworthy that if we just disregarded the self--duality constraint
and we considered the ordinary action of the system composed by the
graviton and an unrestricted 4--form
\begin{equation}
{\cal S}=\frac 1{2 \kappa_{(10)}^2}\int d^{10}x \sqrt{g_{(10)}}\left(R_{(10)} - 
\frac 1{2 \cdot 5!} F_{(5)}^2 \right)
\end{equation}
then, by ordinary variation with respect to the metric, we would anyhow
obtain, as source of the Einstein equation, a traceless
stress--energy tensor:
$$
T_{MN}= \frac 1{2 \cdot 4!}\left(F^2_{(5)MN} - \frac 1{2 \cdot 5} g_{MN} F^2_{(5)}
\right)
$$
The tracelessness of $T_{MN}$ is peculiar to the 4--form and signals its
conformal invariance. This, together with the absence of couplings to the dilaton
(see for instance \cite{bac}), allows for zero curvature solutions in ten dimensions. 

For the metric, we make a block--diagonal ansatz with a Ricci--flat compact part
depending only on the internal coordinates $y^a$ (this corresponds to
choosing the unique Ricci flat
K\"ahler metric on ${\cal M}^{CY}_3$),
and a non--compact part which depends only on the corresponding
non--compact coordinates $x^\mu$
\begin{equation}
ds^2 = g^{(4)}_{\mu\nu}(x)dx^\mu dx^\nu + g^{(6)}_{ab}(y) dy^a dy^b
\end{equation}
For $g^{(4)}_{\mu\nu}$ we  take the extremal R--N black
hole solution, as will be justified below.
This ansatz is consistent with the physical situation under consideration.
In general, the compact components of the metric depend on the non--compact
coordinates
$x^\mu$, being some of the scalars of the $N=2$ effective theory. More
precisely, using complex notation, the components $g_{ij^\star}$ are related
to the $h^{(1,1)}$ moduli parametrizing the deformations of the K\"ahler
class while the $g_{ij}$ ($g_{i^\star j^\star}$) ones are related to the
$h^{(1,2)}$ moduli parametrizing the deformations of the complex structure.
In Type IIB compactifications, as already stressed, such moduli belong to
hypermultiplets and vector multiplets respectively. In our case, however,
there are no vector multiplet scalars, that would couple non--minimally to
the gauge fields (it is usually said that they ``dress'' the field strengths)
and the hypermultiplet scalars can be set  to zero since they do not
couple  to the unique gauge field of our game, namely the graviphoton
(therefore $g_{ab}(x,y) = g_{ab}(y)$).

The 5--form field strength can be generically decomposed in the basis of all
the harmonic 3--forms of the CY manifold $\Omega^{(i,j)}$
\begin{equation}
\label{F5}
F_{(5)}(x,y)=F^0_{(2)}(x) \wedge \Omega^{(3,0)}(y) +
\sum_{k=1}^{h^{(2,1)}} F^k_{(2)}(x) \wedge \Omega_k^{(2,1)}(y) + \mbox{c.c.}
\end{equation}
In the case at hand, however, only the graviphoton $F^0_{(2)}$ appear in the
general ansatz (\ref{F5}), without any additional vector multiplet field
strength $F_{(2)}^k$, and conveniently normalizing
\begin{equation}
\label{F5noi}
F_{(5)}(x,y)= \frac 1{\sqrt{2}} F^0_{(2)}(x) \wedge \left(\Omega^{(3,0)} +
\bar \Omega^{(0,3)}\right)
\end{equation}
Notice that this same ansatz is the consistent one for any double--extreme
solution even for a more generic CY (i.e. with $h^{(1,2)} \neq 0$).

With these ans\"atze, eq. (\ref{E10}) reduces to the usual four--dimensional
Einstein equation with a graviphoton source, the compact part being
identically satisfied. The latter is a non trivial consistency condition
that our ansatz has to fulfil. In fact, in general, eq. (\ref{E10}) taken with
compact indices gives rise (after integration on the compact manifold) to
various equations for the scalar fields. Indeed, the
compact part of the ten--dimensional Ricci tensor $R_{ab}$ is made of the
CY Ricci tensor (that with our choice of the metric is zero by definition)
plus mixed components (i.e. $R^{\mu}_{a\mu b}$) containing, in particular,
kinetic terms of the scalars. The corresponding stress--energy tensor
compact components on the right
hand side of the equation would represent coupling terms of the scalars with
the gauge fields. In our case, however, these  mixed components of $R_{ab}$
are absent. Therefore the complete ten--dimensional Ricci tensor
vanishes ($R_{ab}=0$) and self--consistency of the solution requires that
also the complete stress--energy tensor
$T_{ab}$ should vanish. This follows from our ansatz
(\ref{F5noi}) as it is evident by doing an explicit computation.
This conclusion can also be reached by observing
that the kinetic term of the 4--form does not depend on $g_{ab}$ when $g_{ij}=0$, 
see eq. (\ref{id}) below.

The four--dimensional Lagrangian is obtained by carrying out explicit
integration over the CY. Indeed, choosing the normalization of $\Omega^{(3,0)}$ and
$\bar \Omega^{(0,3)}$ such that $\left\|\Omega^{(3,0)}\right\|^2=V^2_{D3}/V_{CY}$
(since the volume of the corresponding 3--cycle is precisely the volume $V_{D3}$
of the wrapped 3--brane) one has
($z^a = 1/\sqrt{2} (y^a + i y^{a+1})$ and $d^6y = i d^3z d^3\bar z$)
\begin{equation}
\label{id}
\int_{CY} d^6y \sqrt{g_{(6)}} =  V_{CY} \;,\;\; 
i \int_{CY} \Omega^{(3,0)} \wedge \bar \Omega^{(0,3)} =  
V^2_{D3} = \int_{CY} d^6y \sqrt{g_{(6)}} 
\left\|\Omega^{(3,0)}\right\|^2
\end{equation}
and then
\begin{equation}
{\cal S}=\frac 1{2 \kappa_{(4)}^2}\int d^{4}x \sqrt{g_{(4)}}\left(R_{(4)} -
\frac 1{2 \cdot 2!} \mbox{Im}{\cal N}_{00} F^0_{\mu \nu}F^{0|\mu \nu} \right)
\end{equation}
where $\kappa_{(4)}^2 = \kappa_{(10)}^2 / V_{CY}$ and 
$\mbox{Im}{\cal N}_{00}= V_{D3}^2/V_{CY}$. 
In the general case (eq. (\ref{F5})) integration over the CY gives rise, of course,
to a gauge field kinetic term of the standard form: 
$\mbox{Im} {\cal N}_{\Lambda \Sigma} F^{\Lambda} F^{\Sigma}+\mbox{Re} 
{\cal N}_{\Lambda \Sigma}F^{\Lambda} {^*F}^{\Sigma}$, where 
$\Lambda,\Sigma=0,1,...,h^{(1,2)}$.
As well known (from now on $F_{(2)}^0\equiv F$), the four--dimensional Maxwell--Einstein 
equations of motion following from this Lagrangian admit the extremal R--N black hole 
solution (in coordinates in which the horizon is located at $r=0$)
\begin{equation}
\label{RNg}
\begin{array}{cclcccl}
g_{00} \!\!\!&=&\!\!\! 
\displaystyle{-\left(1 + \frac {\kappa_{(4)} M}r \right)^{-2}} &,\;&
g_{mm} \!\!\!&=&\!\!\!
\displaystyle{\left(1 + \frac {\kappa_{(4)} M}r \right)^2} \\ [3mm]
F_{m0} \!\!\!&=&\!\!\! \displaystyle{\kappa_{(4)} \, e_{0} \, \frac {x^m}{r^3} 
\left(1 + \frac {\kappa_{(4)} M}r \right)^{-2}} &,\;& 
F_{mn} \!\!\!&=&\!\!\! 
\displaystyle{\kappa_{(4)} \, g_{0} \, \epsilon_{mnp} \frac {x^p}{r^3}}
\end{array}
\end{equation}
where $m,n,p=1,2,3$. The extremality condition is $M^2 = (e^2 + g^2)/4$, where for 
later convenience we parametrize the solution with
\begin{equation}
\label{char}
M=\frac {\hat \mu}4 \;,\;\; e = e_{0} \sqrt{\frac {V_{D3}^2}{V_{CY}}}=
\frac {\hat \mu}2 \cos \alpha \;,\;\; 
g = g_{0}  \sqrt{\frac {V_{D3}^2}{V_{CY}}}=\frac {\hat \mu}2 \sin \alpha
\end{equation}
The parameter $\hat \mu$ is related to the 3--brane tension $\mu$ through
$\hat \mu = \sqrt{V_{D3}^2/V_{CY}} \mu$, and the arbitrary angle $\alpha$ depends on
the way the 3--brane is wrapped on the CY. Notice that the charges with respect to 
the gauge field $A^\mu$ are $e_0$ and $g_0$, but since the kinetic term, and 
correspondingly the propagator of $A^\mu$, is not canonically normalized, the 
effective couplings appearing in a scattering amplitude are rather $e$ and $g$, 
which indeed satisfy the usual BPS condition. 
Further, at the quantum level, $e$ and $g$ are quantized as a consequence of Dirac's
condition $e g = 2 \pi n$; correspondingly, the angle $\alpha$ can take only discrete
values and this turns out to be automatically implemented in the compactification \cite{bis}.

Now note that in the case of the $T^6/\ZZ_3$ the square volume of the
wrapped $D3$--brane $V^2_{D3}$ defined by the second of eqs. \eqn{id}
is automatically a constant just because the number of vector
multiplets is zero. Notice that for a generic CY compactification we have:
$$
i \, \int_{CY} \Omega^{(3,0)} \wedge \bar \Omega^{(0,3)} \, =
\, \exp \left [ \, {\cal K}\left( \phi , {\bar \phi }\right) \right]
$$
where ${\cal K}\left( \phi , {\bar \phi }\right)$ is the K\"ahler
potential of the moduli fields $\phi(x)$ associated with complex
structure deformations. Hence in the generic case   the $D3$--brane
volume is dressed by scalar fields and depends on the $x$--space
coordinates. Telling the story in four--dimensional language the graviphoton
couples non--minimally to scalar fields. However, on the hand to oxidize the R--N 
type of black--hole solution we discuss in this paper, it is crucial that we
can treat the $D3$--brane square volume $V^2_{D3}$ as $x$--space
independent. 

This ends the field theory side of the computation. Let us turn to a microscopic 
string theory description of the same black--hole.

\vskip 7pt

The problem of describing curved D--branes, such as D--branes wrapped on a
cycle of the internal manifold in a generic compactification of string
theory, is in general too difficult to be solved. In fact, Polchinsky's
description \cite{pol} of D--branes as hypersurfaces on which open strings
can end relies on the possibility of implementing the corresponding boundary
conditions in the CFT describing open string dynamics. Very little has been 
done for a generic target space compactification (for a recent discussion of 
this and related issues, see \cite{gep}) but there exist special cases, 
such as orbifold compactifications, which capture all the essential features of 
more general situations, in which ordinary techniques can be applied.

In previous works \cite{hins1,bis}, a boundary state description of a
D3--brane wrapped on 3--cycle of the $T^6/ \ZZ_3$ orbifold has been proposed
and applied to various situations. In particular, the semiclassical
phase--shift between two of these point--like configurations moving with
constant velocities can be obtained simply by computing the tree level
(cylinder) closed string propagation between the two boundary states
\cite{hins1}. The result is found to vanish like $V^2$ for small relative
velocities, indicating BPS saturation.
The behaviour for large impact parameters, where an effective
description in terms of the underlying low energy four--dimensional
N=2 supergravity is expected to hold, is
\begin{equation}
\label{amp}
{\cal A} = \frac {\hat \mu^2}{4} \left(\cosh v - \cosh 2v \right) 
\int dt \Delta_3(r)
\end{equation}
$v$ being the relative rapidity of the two branes, $\Delta_3(r)$ the 
three--dimensional Green function, $r = \sqrt{b^2 + \sinh^2 v t^2}$ and
$\vec b$ is the impact parameter. In four dimensions, the
exchange of scalar, vector and tensor massless particles between the
two brane sources give contributions with a peculiar dependence on the
rapidity and are proportional to $1$, $\cosh v$ and $\cosh 2v$ respectively.
This leads to the interpretation of eq. (\ref{amp}) as the exchange of the
bosonic part of the N=2 gravitational multiplet, that is the graviton and
the graviphoton. The absence of any constant part in (\ref{amp}) signals
that there is no scalar exchange between the two branes. Since the two branes
are identical and therefore have the same coupling to the scalars of the bulk
four--dimensional supergravity, the total scalar exchange is proportional to the
sum of the squares of these couplings, and its vanishing implies the vanishing
of all the couplings separately. It is interesting to compare (\ref{amp}) to
the result for a 0--brane (arising in a corresponding IIA compactification)
\begin{equation}
\label{amp0}
{\cal A} = \frac {\hat \mu^2}{4} \left(4 \cosh v - \cosh 2v - 3 \right)
\int dt \Delta_3(r)
\end{equation}
for which scalars are exchanged, beside the graviton and the vector. Since the 
ten--dimensional 0-brane couples only to the dilaton $\phi^{(10)}$ and the 
world-volume components of the graviton $h^{(10)}_{\mu \nu}$ and the RR
vector $A^{(10)}_\mu$, the four--dimensional 0--brane couples only to the
corresponding four--dimensional fields $\phi^{(4)}$, $h^{(4)}_{\mu \nu}$
and $A^{(4)}_\mu$ (in particular, in the four--dimensional Einstein frame,
it does not couple to the additional scalars and vectors coming from metric). 

For the wrapped 3--brane, eq. (\ref{amp}), BPS saturation implies that all 
the vector repulsion is balanced only by gravitational attraction, whereas for
the 0--brane, also the scalars contribute to the attraction, leaving a smaller
gravitational potential. Actually both of these four--dimensional 
configurations come from an effective action of the type
\begin{equation}
{\cal S}=\int d^{4}x \sqrt{g}\left(R -
\frac 12 \left(\partial \phi \right)^2 -\frac 1{2 \cdot 2!} 
e^{- a \phi} F_{(2)}^2 \right)
\end{equation} 
with $a=0$ for the R--N black hole and $a \neq 0$ for the 0--brane. The general 
electric extremal solution of this Lagrangian is \cite{stelle}
\begin{equation}
ds^2 = - H(r)^{-\alpha} dt^2 + H(r)^{\alpha} d \vec x \cdot d \vec x \;,\;\;
\phi = \beta \, \ln H(r) \;,\;\;
A_0 = \gamma \, H(r)^{-1}
\end{equation}
where
\begin{equation}
\alpha = \frac 2{1 + a^2} \;,\;\; \beta = \frac {2a}{1 + a^2} \;,\;\; 
\gamma = \frac {2}{\sqrt{1 + a^2}} 
\end{equation}
and $H(r)$ satisfies the three-dimensional Laplace equation and can be taken to be 
of the form $H(r) = 1 + k \Delta_3(r)$.
The relevant asymptotic long range fields are thus
$$
h_{00} = \alpha \, k \, \Delta_3(r) \;,\;\; \phi = \beta \, k \, \Delta_3(r)
\;,\;\; A_0 = \gamma \, k \, \Delta_3(r)
$$
and so the phase-shift between two identical branes moving with relative
rapidity $v$ is
\begin{equation}
{\cal A} = k^2 \left(\gamma^2 \cosh v - \alpha^2 \cosh 2v - \beta^2 \right) 
\int dt \Delta_3(r)
\end{equation}
As a consequence of BPS saturation, $\beta^2 - \alpha^2 - \gamma^2 = 0$ and the
static force vanishes. Moreover, comparing with eqs. (\ref{amp}) and (\ref{amp0}),
we learn that the R--N solution corresponds to $a=0$ and $k=\hat \mu / 4$, whereas the
0--brane corresponds to $a=\sqrt{3}$ and $k=\hat \mu$.

Altogether, these arguments lead to evidence that the boundary state constructed in 
refs \cite{hins1,bis} actually represents a R--N black hole. An equivalent way of 
analyzing this configuration, to see again that it indeed correctly fits the general 
solution R--N $\times$ CY discussed before, is to compute one--point functions 
$\langle\Psi\rangle=\langle\Psi|B\rangle$ of the massless fields of 
supergravity and compare them with the linearized long range fields of the 
supergravity R--N black hole solution (\ref{RNg}). This second method presents the
advantage of yielding direct informations on the coulpings with the massless fields
of the low energy theory.

The original ten--dimensional coordinates are organized as follows: the
four non--compact directions $X^0,X^1,X^2,X^3$ span ${\cal M}_4$, whereas
the six compact directions $X^a,X^{a+1}$, $a=4,6,8$, span $T^6/ \ZZ_3$.
The three $T^2$'s composing $T^6$ are parametrized by the 3 pairs
$X^a,X^{a+1}$, and the $\ZZ_3$ action is generated by $2\pi/3$ rotations
in these planes. The boundary state $|B\rangle$ of the D3--brane wrapped
on a generic $\ZZ_3$--invariant 3--cycle can be obtained from the boundary
state $|B_3(\theta_0)\rangle$ of D3--brane in ten dimensions with Neumann
directions $X^0$ and $X^{\prime a}(\theta_0)$, where the $X^{\prime a}(\theta_0)$
directions form an arbitrary common angle $\theta_0$ with the $X^{a}$ directions in
each of the 3 planes $X^a,X^{a+1}$ (actually, we could have chosen 3 different angles
in the 3 planes, but only their sum will be relevant, as it could be inferred 
from eq. (\ref{proj}) below). First, one projects onto the
$\ZZ_3$--invariant part and then compactifies the directions $X^a,X^{a+1}$.
The $\ZZ_3$ projection is implemented by applying the projector
$P= 1/3 (1 + g + g^2)$ on $|B_3(\theta_0)\rangle$, where
$g=\exp i 2\pi/3 (J^{45} + J^{67} + J^{89})$ is the generator of the
$\ZZ_3$ action and $J^{aa+1}$ is the $X^a,X^{a+1}$ component of the angular momentum
operator. This yields
\begin{equation}
\label{bo}
|B\rangle = \frac 13 \sum_{\{\Delta \theta\}}
|B_3(\theta = \Delta \theta + \theta_0)\rangle
\end{equation}
where the sum is over $\Delta \theta = 0, 2\pi/3,4\pi/3$.
It is obvious form this formula that $|B\rangle$ is a periodic function of the
parameter $\theta_0$ with period $2\pi/3$. Therefore, the physically distinct
values of $\theta_0$ are in $[0,2\pi/3]$ and define a one parameter family of
$\ZZ_3$--invariant boundary states, corresponding to all the possible harmonic 
3--forms on $T^6/ \ZZ_3$, as we will see.
Notice that requiring a fixed finite volume $V_{D3}$ for the 3--cycle on which
the D3--brane is wrapped implies discrete values for $\theta_0$ \cite{bis}.
The compactification process restricts the
momenta entering the Fourier decomposition of $|B\rangle$ to belong the
momentum lattice of $T^6/ \ZZ_3$. Since the massless supergraviton
states $|\Psi\rangle$ carry only space time momentum, the compact part of
the boundary state will contribute a volume factor which turns the 
ten--dimensional D3--brane tension $\mu=\sqrt{2\pi}$ into the four--dimensional 
black hole charge $\hat \mu = \sqrt{V_{D3}^2/V_{CY}} \mu$ \cite{bis}, and some 
trigonometric functions of $\theta_0$ to be discussed below.

Using the technique of ref. \cite{torino}, the relevant one--point functions
on $|B_3(\theta)\rangle$ for the graviton and 4--form states $|h\rangle$ and
$|A\rangle$ with polarization $h^{MN}$ and $A^{MNPQ}$, are 
\begin{equation}
\label{htt}
\langle B_3(\theta)|h\rangle = - \frac {\hat \mu}{2} \,  T \,
h_{MN}\,  M^{MN}(\theta) \;,\;\;
\langle B_3(\theta)|A\rangle = - \frac {\hat \mu}{8}\,  T \,
A_{MNPQ} \, M_{ab}(\theta)\,  \Gamma^{MNPQ}_{ba}
\end{equation}
$T$ is the total time and $\mu$ is correctly changed to $\hat \mu$ by the volume
factor that the compact part of the boundary state contributes \cite{bis}.
The numerical coefficients appearing in (\ref{htt}) have been choosen
at our convenience by relying on the scattering amplitude \cite{bis}, where the 
relative normalization is easily fixed, as already discussed.
The matrices $M(\theta)=\Sigma(\theta) M \Sigma^T(\theta)$ are obtained
from the usual ones corresponding to Neumann boundary conditions along 
$X^0,X^4,X^6,X^8$
$$
M_{MN} = \mbox{diag}(-1,-1,-1,-1,1,-1,1,-1,1,-1) \;,\;\;
M_{ab} = \Gamma^{0468}_{ab}
$$
through a rotation of angle $\theta$ in the 3 planes $X^a,X^{a+1}$, generated
in the vector and spinor representations of each $S0(2)$ subgroup of the rotation
group $S0(8)$ by
$$
\Sigma_V(\theta)=\pmatrix {\cos \theta & \sin \theta \cr -
\sin \theta & \cos \theta \cr}\;,\;\;
\Sigma_S(\theta)=\cos \frac{\theta}{2} - \sin \frac{\theta}{2} \, \Gamma^{aa+1}
$$
After some simple algebra, one finds 
\begin{eqnarray}
\label{ht}
\spa \langle B_3(\theta)|h\rangle = \frac {\hat \mu}{ 2} \,   T \,
\left\{h^{00}+h^{11}+h^{22}+h^{33} 
- \sum_a \left[\cos 2 \theta \, \left( h^{aa} - h^{a+1a+1}\right)
- 2 \sin 2 \theta \, h^{aa+1} \right] \right\} \nn \\ 
\spa \langle B_3(\theta)|A\rangle = 2 \hat \mu T
\left[\cos^3 \theta \left(A^{0468} - A^{0479} - A^{0569} - A^{0578}\right)
\right. \nn \\ \spa \qquad \qquad \qquad \qquad 
+ \sin^3 \theta \left(A^{0579} - A^{0568} - A^{0478} - A^{0469} \right) \nn
\\ \spa \qquad \qquad \qquad \qquad \left.
+ \cos \theta \left(A^{0479}+A^{0569}+A^{0578}\right)
+ \sin \theta \left(A^{0568}+A^{0478}+A^{0469}\right) \right]
\end{eqnarray}

The one--point functions for the $D3$--brane wrapped on $T^6/ \ZZ_3$ are then obtained
by averaging over the allowed $\Delta \theta$'s:
$\langle\Psi\rangle = 1/3 \sum_{\{\Delta \theta\}} \langle B_3(\theta)|\Psi\rangle$.
One easily finds the only non--vanishing averages of the trigonometric
functions appearing in eq.s \eqn{ht} to be
\begin{equation}
\label{proj}
\frac 13 \sum_{\{\Delta \theta\}} \cos^3 \theta =
\frac 14 \cos 3 \theta_0 \;,\;\;
\frac 13 \sum_{\{\Delta \theta\}} \sin^3 \theta =
- \frac 14 \sin 3 \theta_0
\end{equation}
so that finally, meaning now with $h$ and $A$ all the four--dimensional 
fields arising from the graviton and the 4--form respectively 
upon compactification,
\begin{equation}
\label{h}
\langle h\rangle = \frac {\hat \mu}2 T
\left(h^{00}+h^{11}+h^{22}+h^{33}\right) \;,\;\;
\langle A\rangle =  \frac {\hat \mu}2 T \left(
\cos 3 \theta_0 A^0 - \sin 3 \theta_0 B^0 \right)
\end{equation}
where we have defined the graviphoton fields
\begin{equation}
A^\mu \equiv A^{\mu 468} - A^{\mu 479} - A^{\mu 569} - A^{\mu 578}
 \;,\;\; B^\mu \equiv
A^{\mu 579} - A^{\mu 568} - A^{\mu 478} - A^{\mu 469}
\end{equation}
Using self--duality of the 5--form field strength in ten dimension, one easily see that
$F_B^{\mu\nu} = {^*F}_A^{\mu\nu}$
so that $A^\mu$ and $B^\mu$ are not independent fields, but rather magnetically dual.
Using the $A^\mu$ field, we get electric and magnetic charges
\begin{equation}
\label{chars}
e = \frac {\hat \mu}2 \cos 3 \theta_0 \;,\;\; g = \frac {\hat \mu}2 \sin 3 \theta_0
\end{equation}
or viceversa using the $B^\mu$ field. Comparing with eq. (\ref{char}) one finds that
$\alpha = 3 \theta_0$ and therefore the ratio between $e$ and $g$ depends on the 
choice of the 3--cycle, as anticipated. Also, as explained, only discrete values of
$\theta_0$ naturally emerge requiring a finite volume. 

Further evidence for the identifications \eqn{chars}  comes from the computation
of the electromagnetic phase--shift between two of these configurations with different
$\theta_0$'s, call them $\theta_{1,2}$.
Since the four--dimensional electric and magnetic charges of the
two black holes are then different, there should be both an even and an odd
contribution to the phase--shift coming from the corresponding R--R spin structures.
Indeed, one correctly finds \cite{bis}
\begin{equation}
{\cal A}_{even} \sim \frac {{\hat \mu}^2}4 
\cos 3 \left(\theta_{1} - \theta_{2} \right) =
e_1 e_2 + g_1 g_2 \;,\;\; {\cal A}_{odd} \sim 
\frac {{\hat \mu}^2}4 \sin 3 \left(\theta_{1} - \theta_{2} \right) =
e_1 g_2 - g_1 e_2
\end{equation}

Notice that all the compact components $h^{ab}$ of the graviton have cancelled
in (\ref{h}),
reflecting the fact the black hole has no scalar hairs. Moreover, the one--point
function of the R--R 4--form is precisely of the form of our ansatz (\ref{F5noi}),
with the unique holomorphic and antiholomorphic 3--forms $\Omega^{(3,0)}$ and
$\bar \Omega^{(0,3)}$ showing up in (\ref{h}). Indeed
\begin{equation}
\label{obo}
\Omega^{(3,0)} = \Omega \, dz^4 \wedge dz^6 \wedge dz^8\;,\;\;
\bar \Omega^{(0,3)} = \Omega^* \, d\bar z^4 \wedge d\bar z^6 \wedge d\bar z^8
\end{equation}
so that the real 3--form appearing in (\ref{F5noi}) is given by
\begin{equation}
\label{3003}
\Omega^{(3,0)} + \bar \Omega^{(0,3)} = \mbox{Re}\Omega
\left(\omega^{468} - \omega^{479} - \omega^{569} - \omega^{578}\right)
+\mbox{Im}\Omega
\left(\omega^{579} - \omega^{568} - \omega^{478} - \omega^{469}\right)
\end{equation}
where $\omega^{abc} = 1/\sqrt{2} \, dy^a \wedge dy^b \wedge dy^c$.
The precise correspondence between the boundary state result (\ref{h}) and the purely
geometric identity (\ref{3003}) is then evident.
The combination of components of the 4--form appearing in (\ref{h}) is
proportional to the integral over the D3--brane world--volume $V_{1+3}$
\begin{equation}
\label{A4A0}
\langle A\rangle = \frac {\mu}2 \, \mbox{Re} \int_{V_{1+3}} \left(A + i B \right) \wedge
\Omega^{(3,0)} = \int_{V_1} \left(e A + g B \right)
\end{equation}
This formula yields an interesting relation between the parameters
$\mu, \hat \mu,\theta_0$ and the complex component $\Omega$ in (\ref{obo}) defining
the 3--cycle; one gets $\Omega = (\hat \mu / \mu) e^{-i 3 \theta_0}$.
Notice that one correctly recovers $| \Omega |= \sqrt{V^2_{D3}/V_{CY}}$,
the arbitrary phase being the sum of the arbitrary overall angles $\theta_0$
appearing in the boundary state construction.
Finally, dropping the overall time $T$, inserting a propagator $\Delta = 1/\vec q^2$
and Fourier transforming eqs. (\ref{h}) with the identification
(\ref{A4A0}), one recovers the asymptotic gravitational and electromagnetic fields
of the R--N black hole, eqs. (\ref{RNg}).

This confirms that our boundary state describes a $D3$--brane
wrapped on $T^6/ \ZZ_3$, falling in the class of regular four--dimensional
R--N double--extreme black holes obtained by wrapping the self--dual
D3--brane on a generic CY threefold. This boundary state encodes the
leading order couplings to the massless fields of the theory, and allows the
direct determination of their long range components, falling off like $1/r$ in
four dimensions. The subleading post--Newtonian corrections to these fields
arise instead as open string higher loop corrections, corresponding to string
world--sheets with more boundaries; from a classical field theory point of view, 
this is the standard replica of the source in the tree-level perturbative evaluation 
of a non--linear classical theory. In a series expansion for $r \rightarrow \infty$, a
generic term going like $1/r^{l}$ comes from a diagram with $l$ open string
loops, that is $l$ branches of a tree-level closed string graph (each branch 
brings an integration over the transverse 3--momentum, two
propagators and a supergravity vertex involving two powers of momentum, yielding
an overall contribution of dimension $1/r$).

\vskip 7pt

Let us end with few final comments. As pointed out by the authors of
\cite{lust1}, heuristically speaking the reason why single D--brane black holes
are non--singular in CY compactifications, as opposed to the toroidal
case, is that the brane is wrapped on a topologically non--trivial
manifold and therefore can intersect with itself. This intersection mimics
the actual intersecting picture of different branes holding in toroidal
compactifications that is the essential feature in order to get a non--singular
solution in that case. In our case, such analogy is particularly manifest
since the boundary state $\ZZ_3$--invariant projection (\ref{bo}) can be seen as
a three D3--branes superposition at angles ($2 \pi/3$) in a $T^6$ compactification.
As illustrated in \cite{tow} such intersection preserves precisely $1/8$
supersymmetry, as a single D3--brane does on $T^6/ \ZZ_3$. For toroidal
compactification this is not enough, of course, because at least 4
intersecting D3--branes are needed in order to get a regular solution \cite{bal}.

Finally, since this extremal R--N configuration is constructed by a single D3--brane, 
it naturally arises the question of understanding the microscopic origin of its entropy.

\vskip 10pt

{\bf \large Acknowledgments}

M. B. and C. A. S. are grateful to G. Bonelli for enlighten discussions and to the
Dipartimento di Fisica Teorica of Torino University for hospitality.

\end{document}